\documentclass{emulateapj}

\shorttitle{2D RHD MODEL FOR LIMIT-CYCLE DISK OSCILLATIONS}
\shortauthors{OHSUGA}

\newcommand{\lsim}{\raisebox{0.3mm}{\em $\, <$} \hspace{-2.8mm}
\raisebox{-1.8mm}{\em $\sim \,$}}
\newcommand{\gsim}{\raisebox{0.3mm}{\em $\, >$} \hspace{-2.8mm}
\raisebox{-1.8mm}{\em $\sim \,$}}
\newcommand{\bm}[1]{\mbox{\boldmath $#1$}}

\begin{document}

\title{Two-dimensional radiation-hydrodynamic model for 
limit-cycle oscillations of luminous accretion disks}


\author{K. Ohsuga}
\affil{Department of Physics, Rikkyo University, 
Toshimaku, Tokyo 171-8501, Japan}

\begin{abstract}
We investigate the time evolution of 
luminous accretion disks around black holes,
conducting the two-dimensional radiation-hydrodynamic simulations.
We adopt the $\alpha$ prescription for the viscosity.
The radial-azimuthal component of viscous stress tensor
is assumed to be proportional to the total pressure 
in the optically thick region,
while the gas pressure in the optically thin regime.
The viscosity parameter, $\alpha$, is taken to be 0.1.
We find the limit-cycle variation in luminosity
between high and low states.
When we set the mass input rate from the outer disk boundary 
to be $100 L_{\rm E}/c^2$,
the luminosity suddenly rises
from $0.3L_{\rm E}$ to $2L_{\rm E}$,
where $L_{\rm E}$ is the Eddington luminosity.
It decays after retaining high value for about $40$ s.
Our numerical results 
can explain the variation amplitude and duration 
of the recurrent outbursts observed in microquasar,
GRS 1915+105.
We show that the multi-dimensional effects 
play an important role in the high-luminosity state.
In this state, 
the outflow is driven by the strong radiation force,
and some part of radiation energy dissipated inside the disk
is swallowed by the black hole due to the photon-trapping effects.
This trapped luminosity is comparable to the disk luminosity.
We also calculate two more cases: one with a much larger accretion rate 
than the critical value for the instability and the other with the 
viscous stress tensor being proportional to the gas pressure only
even when the radiation pressure is dominant.
We find no quasi-periodic light variations in these cases. 
This confirms that the 
limit-cycle behavior found in the simulations is caused by the disk 
instability.
\end{abstract}


\section{INTRODUCTION}
Microquasars in our Galaxy display large flux variability 
in X-ray band.
Such luminosity variations are thought to reflect violent phenomena 
in accretion disks around black holes.
The dwarf-nova type disk-instability
is responsible for the luminosity changes on the time-scale of months
(e.g., Mineshige \& Wheeler 1989; for a review Kato et al. 1998, \S 5). 
In contrast, the mechanism causing the short-term variability 
($10-100$ s)
is not well understood yet.
One of the plausible mechanisms for it 
is thermal and secular instability 
which arises when the radiation pressure becomes dominant over gas pressure
(e.g., Lightman \& Eardley 1974; Shibazaki \& H\=oshi 1975;
Pringle 1976; Shakura \& Sunyaev 1976).
S-shaped sequence on the $\dot{M}_{\rm acc}-\Sigma$ 
(mass accretion rate vs. surface density) plane
was completed by the finding of a slim-disk branch (Abramowicz et al. 1998),
in which the viscous heating is balanced by the advective cooling.
It was suggested that the disk may undergo limit-cycle oscillations,
like the dwarf nova outbursts
(for a review Kato et al. 1998, \S 10).
This instability is expected to occur when
the mass accretion rate is comparable to or 
moderately exceeds the critical value,
$L_{\rm E}/c^2$, 
where $L_{\rm E}$ is the Eddington luminosity and 
$c$ is the velocity of light.

The limit-cycle oscillations have been investigated 
by one-dimensional (1D), vertically-integrated approach. 
The quasi-periodic outbursts were
first demonstrated by Honma et al. (1991) 
using the time-dependent 1D simulations of the accretion disks
and have then been investigated in detail 
(Szuszkiewicz \& Miller 1997, 1998, 2001; 
Watarai \& Mineshige 2003).
However, 
the 1D model cannot treat the multi-dimensional motion, 
i.e., convection, circulation, and outflow,
although they would influence the flow dynamics significantly
via the transport of mass, momentum/angular momentum, 
and energy.
The modification of the accretion disk model,
where the mass ejection from the disk surface is taken into account,
was suggested by Nayakshin et al. (2000) 
(see also Janiuk et al. 2000; 2002 and Janiuk \& Czerny 2005).
But they still used simplified 1D method without 
treating the multi-dimensional flow motion.
We need to perform at least two-dimensional (2D) analysis
in order to understand the accretion flow correctly.

Recently, Teresi et al. (2004a, 2004b) first
reproduced the quasi-periodic luminosity variations
by a 2D smoothed particle hydrodynamic (SPH) simulations.
However, their simulations focused on the time evolution of the disk itself
and do not treat the optically thin regime correctly, i.e.,
the disk surface as well as the atmosphere surrounding the disk.
They assumed the equilibrium between gas and radiation 
without treating the energy of gas and radiation separately.
In their simulations, 
the transport of the radiation energy was not solved 
in the optically thin regime,
though the radiation flux was evaluated by 
using the diffusion approximation deep inside the disk.
The thermal energy was extracted 
from the SPH particles located near the disk surface.
It should be calculated by solving the interaction 
between gas and radiation as well as radiative transfer.
Their assumptions and method are valid only in the optically thick regime,
i.e., deep inside the disk.
As a result, the outflow was not found in their simulations
even though the luminosity exceeded the Eddington luminosity
in the high-luminosity state.
The outflow would have an important role on the evolution of the disk
through the extraction of the mass, momentum/angular momentum, 
and energy from the disk.
In addition, the simulations for stable disks were not performed in 
their work.
A comparison of their results with those of simulations for 
stable disks is essential
in order to understand the physical mechanism of the disk oscillations.
Hence, it is important to confirm the limit-cycle behavior
by the grid-based simulations,
in which the multi-dimensional radiation hydrodynamic (RHD) equations
are properly solved.
It is also important to investigate the time evolution of 
not only an unstable disk but also a stable disk.

The photon trapping is also basically a multi-dimensional effect,
by which some or large part of photons generated inside the disk is swallowed 
by the black hole
in supercritical accretion regime, 
leading to a reduction of the energy conversion efficiency
(Ohsuga et al. 2002, 2003, 2005).
It has been reported by the analysis of the X-ray data of GRS 1915+105
that the disk luminosity is comparable to or exceeds 
the Eddington luminosity in the high-luminosity state
(Yamaoka et al. 2001), implying the supercritical accretion.
Thus, the photon trapping is expected to appear
in high-luminosity state of the quasi-periodic oscillations.


Here, by solving full set of 2D RHD
equations including viscosity term,
we report the 2D RHD model 
for the quasi-periodic oscillations of the accretion disks 
around black holes.
Our numerical simulations carefully treat the 2D effects,
including the outflow motion and the photon trapping.
We also show in our simulations that the disk is unstable
on condition that 
the mass accretion rate is moderately larger than the critical value 
and the viscous stress tensor is proportional to the total pressure.
It is consistent with the disk theory and
proves that the bursting behavior in our simulations
arises from the disk instability in the radiation-pressure dominant region.
%
%
In \S 2, our model and numerical method are described. 
We present the numerical results in \S 3.
\S 4 and \S5 are devoted to discussion and conclusions.

\section{MODEL AND NUMERICAL METHOD}
Basic equations and our numerical method are described 
in detail in Ohsuga et al. (2005).
Here we briefly summarize the model and the numerical method.
The set of RHD equations including the viscosity term
are solved by 
an explicit-implicit finite difference scheme on the Eulerian grids.
We use spherical polar coordinates $(r, \theta, \varphi)$,
where $r$ is the radial distance, 
$\theta$ is the polar angle,
and $\varphi$ is the azimuthal angle.
In the present study,
we assume the axisymmetry as well as the reflection symmetry
relative to the equatorial plane.
We also assume that 
only $r\varphi$-component of the viscous stress tensor,
$\tau_{r\varphi}$,
is non zero.
This component plays an important role for
the transport of the angular momentum
and heating of the disk plasma.
We describe the gravitational field of the black hole in terms of 
pseudo-Newtonian hydrodynamics (Paczynski \& Wiita 1980), 
and assume the flow to be non self-gravitating.
The basic equations are the continuity equation,
\begin{equation}
  \frac{\partial \rho}{\partial t}
  + {\bm \nabla} \cdot (\rho {\bm v}) = 0,
\end{equation}
the equations of motion,
\begin{eqnarray}
  \frac{\partial (\rho v_r)}{\partial t}
  + {\bm \nabla} \cdot (\rho v_r {\bm v}) 
  &=& - \frac{\partial p}{\partial r} 
  + \rho \left[ 
    \frac{v_\theta^2}{r} + \frac{v_\varphi^2}{r}
    -\frac{GM}{(r-r_s)^2}
  \right] \nonumber\\
  & &+ \frac {\chi}{c} F_0^r,
\end{eqnarray}
\begin{equation}
  \frac{\partial (\rho r v_\theta)}{\partial t}
  + {\bm \nabla} \cdot (\rho r v_\theta {\bm v}) 
  = - \frac{\partial p}{\partial \theta}
  + \rho v_\varphi^2 \cot\theta
  + \frac {\chi}{c} rF_0^\theta,
\end{equation}
\begin{eqnarray}
  \frac{\partial (\rho r \sin\theta v_\varphi)}{\partial t}
  &+& {\bm \nabla} \cdot (\rho r \sin\theta v_\varphi {\bm v}) 
  \nonumber\\
  &=& \frac{1}{r^2}\frac{\partial}{\partial r}
  \left( r^3 \sin\theta \tau_{r\varphi} \right),
\end{eqnarray}
the energy equation of the gas,
\begin{eqnarray}
  \frac{\partial e }{\partial t}
  + {\bm \nabla}\cdot(e {\bm v}) 
  =& & -p_{\rm gas}{\bm \nabla}\cdot{\bm v} -4\pi \kappa B 
  + c\kappa  E_0 \nonumber\\
  & &+\eta 
  \left[ r \frac{\partial}{\partial r}
  \left(\frac{v_\varphi}{r}\right)
  \right]^2,
\end{eqnarray}
and the energy equation of the radiation,
\begin{equation}
  \frac{\partial E_0}{\partial t}
  + {\bm \nabla}\cdot(E_0 {\bm v}) 
  = -{\bm \nabla}\cdot{\bm F_0} -{\bm \nabla}{\bm v}:{\bm {\rm P}_0}
  + 4\pi \kappa B - c\kappa E_0.
  \label{rade}
\end{equation}
Here, $\rho$ is the gas mass density,
$\bm{v}=(v_r, v_\theta, v_\varphi)$ is the velocity,
$p_{\rm gas}$ is the gas pressure,
$e$ is the internal energy density of the gas,
$B$ is the blackbody intensity,
$E_0$ is the radiation energy density,
${\bm F}_0=\left(F_0^r, F_0^\theta \right)$ is the radiation flux,
${\bm {\rm P}}_0$ is the radiation pressure tensor,
$\eta$ is the dynamical viscosity coefficient,
$\kappa$ is the absorption opacity,
and $\chi (= \kappa+\rho \sigma_{\rm T}/m_{\rm p})$ 
is the total opacity 
with $\sigma_{\rm T}$ being the Thomson scattering cross-section
and $m_{\rm p}$ being the proton mass.

For the equation of state we use
$p_{\rm gas}=(\gamma-1)e$,
where $\gamma$ is the specific heat ratio.
The temperature of the gas, $T$, can then be calculated from
$p_{\rm gas}=\rho k_{\rm B} T/\mu m_{\rm p}$,
where $k_{\rm B}$ is the Boltzmann constant and
$\mu$ is the mean molecular weight.
%
To complete the set of equations, we apply
flux limited diffusion (FLD) approximation
developed by Levermore and Pomraning (1981).
Adopting this approximation, the radiation flux and 
the radiation pressure tensor are expressed 
in terms of the radiation energy density
(Turner \& Stone 2001, Ohsuga et al. 2005).

In this paper, the dynamical viscosity coefficient is 
given by
\begin{equation}
  \eta = \alpha \frac{p_{\rm gas}+\lambda E_0}{\Omega_{\rm K}},
\end{equation}
where 
$\alpha$ is the viscosity parameter,
$\Omega_{\rm K}$ is the Keplerian angular speed,
and $\lambda$ is the flux limiter.
In this form, 
the $r\varphi$-component of the viscous stress tensor is 
\begin{equation}
  \tau_{r\varphi} \propto \left\{
    \begin{array}{ll}
      \alpha p_{\rm total} & \mbox{(optically thick limit)} \\
      \alpha p_{\rm gas} & \mbox{(optically thin limit)}
    \end{array}
  \right.,
\end{equation}
where $p_{\rm total}$ is the total pressure.
This is because 
the flux limiter, $\lambda$, becomes $1/3$ 
in the optically thick limit,
and vanishes in the optically thin limit
in the framework of the FLD approximation.
The flux limiter, $\lambda$, is almost $1/3$ in the disk region.
Thus, our viscosity model is basically 
the same as the $\alpha$ prescription of the viscosity
proposed by Shakura \& Sunyaev (1973),
although $\lambda$ is almost null above and below the disk.

The computational domain consists of spherical shells of 
$3r_g \leq r \leq 500r_g$ and $0 \leq \theta \leq 0.5\pi$,
and is divided into $96\times 96$ grid cells,
where $r_g$ is the Schwarzschild radius.
We start the calculations with a hot, rarefied, and 
optically-thin atmosphere. 
There is no cold dense disk initially,
and we assume that matter
is continuously injected into the computational domain
through the outer-disk boundary 
($r=500r_g$, $0.45\pi \leq \theta \leq 0.5\pi$).
Therefore, 
we can avoid the influence of the initial configuration on the final result,
although a long integration time is required.
The injected matter is supposed to 
have a specific angular momentum corresponding to the 
Keplerian angular momentum at $r=100r_g$,
and we set the injected mass-accretion rate (mass input rate)
so as to be constant at the boundary.

Throughout the present study, 
we assume the black-hole mass to be $M=10M_\sun$,
$\alpha=0.1$, $\gamma=5/3$, and $\mu=0.5$.
For the absorption opacity,
we consider the free-free absorption 
and the bound-free absorption for solar metallicity
(Hayashi, Hoshi, \& Sugimito 1962, Rybicki \& Lightman 1979).

\section{RESULTS}
Figure 1 represents the time evolution 
of the mass accretion rate onto the black hole, $\dot{M}_{\rm acc}$ (blue),
the outflow rate, $\dot{M}_{\rm out}$ (magenta),
the luminosity, $L$ (red),
and the trapped luminosity, $L_{\rm trap}$ (green).
Here, 
the outflow rate means the ejected mass per unit time at the 
outer boundary of the computational domain.
The luminosity is evaluated by integrating the radiation flux
at the outer boundary.
On the other hand, the trapped luminosity is the integration
of the radiation flux at the inner boundary,
which means the radiation energy swallowed by the black hole per unit time.
The luminosity, $L$, almost equals to the luminosity of the disk,
since the emission of the less dense atmosphere surrounding the disk
does not contribute very much to the luminosity.
The mass input rate is set to be $\dot{M}_{\rm input}=100 L_{\rm E}/c^2$.

\begin{figure}[b]
\epsscale{1.18}
\plotone{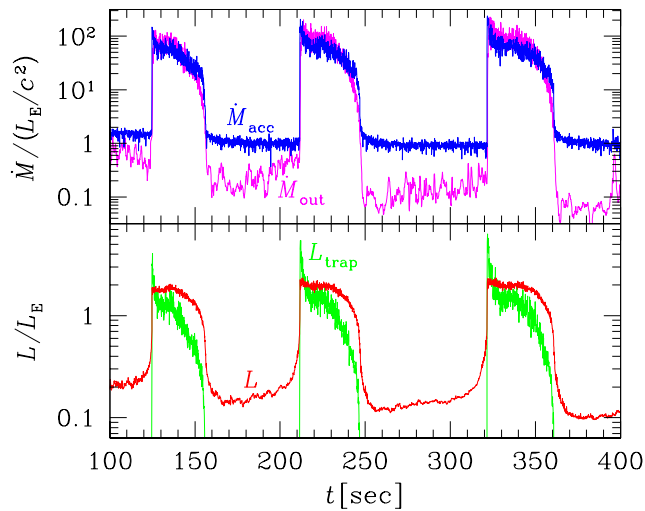}
\caption{
The time evolution of the mass accretion rate (blue), outflow rate (magenta),
the luminosity (red), and trapped luminosity (green),
for $\dot{M}_{\rm input}=100 L_{\rm E}/c^2$.
Here, we employ the viscosity model, which gives 
the simple $\alpha$ prescription
proposed by Shakura \& Sunyaev (1973), 
$t_{r\varphi} \propto p_{\rm total}$, inside the disk.
}
\end{figure}

The injected matter accumulates within the computational domain
and the accretion disk forms around the black hole ($t<100$ s).
Subsequently, the limit-cycle oscillation starts.
At $t>100$s, 
the mass accretion rate onto the black hole ($\dot{M}_{\rm acc}$) 
drastically varies,
although the mass is continuously injected into the computational domain 
at constant rate.
It suddenly rises, retains high value for about 40 s, 
and then decays. 
Such time variation of the mass accretion rate 
causes the quasi-periodic luminosity changes
between high- and low-luminosity states. 
Whereas the luminosity ($L$) is not more than $0.3L_{\rm E}$
in the low-luminosity state, 
it reaches around $2L_{\rm E}$ in the high-luminosity state. 
The burst duration is also about 40 s, which is 
roughly correspond to the viscous timescale,
$ t_{\rm vis}\sim 35.8 (M/10M_\odot)
(r/100r_g)^{3/2} (\alpha/0.1)^{-1} (H/0.2r)^{-2}$ s.
The physical mechanism of such limit-cycle oscillations 
is the thermal instability 
in the radiation-pressure dominant region
(discussed later). 

Moreover, 
we can see in this figure that 
our simulations reveal two important phenomena.
One of them is that the outburst is accompanied 
by the strong outflow. 
As shown in the top panel, while the outflow rate ($\dot{M}_{\rm out}$) 
is very small in the low-luminosity state, 
it suddenly rises together with mass accretion rate 
and retains the high value during the outburst. 
Since the luminosity is larger than the Eddington luminosity,
the outflow is driven by the strong radiation force.
Another one is that the photon-trapping effects appear during the burst. 
It is found that the trapped luminosity ($L_{\rm trap}$) 
is comparable to the luminosity
in the high-luminosity state. 
Thus, 
considerable number of photons falls onto the black hole, 
reducing the apparent luminosity,
although the disk is very luminous in this state. 
In contrast, 
the photon trapping does not work effectively in the low-luminosity state.
The multi-dimensional effects are significant
in the high-luminosity state.

To understand the difference of the disk structure between two states,
we display the cross-sectional view 
of the density distribution in the low-(left panel) and high-luminosity
states (right panel) in Figure 2.
The velocity vectors are overlaid in the right panel.
It is clear that the disk becomes geometrically thick 
and is accompanied by the outflow 
in the high-luminosity state, whereas the thin disk forms 
in the low-luminosity state. 
Further, the patchy structure as well as the circular motion
appears within the disk in the high-luminosity state.
They would be caused by the Kelvin-Helmholtz instability 
around the disk surface and by probably convection 
(see Ohsuga et al. 2005). 
It is stressed again that 
the multi-dimensional effects are significant
in the high-luminosity state.
%
%
\begin{figure*}[t]
\epsscale{1.17}
\plotone{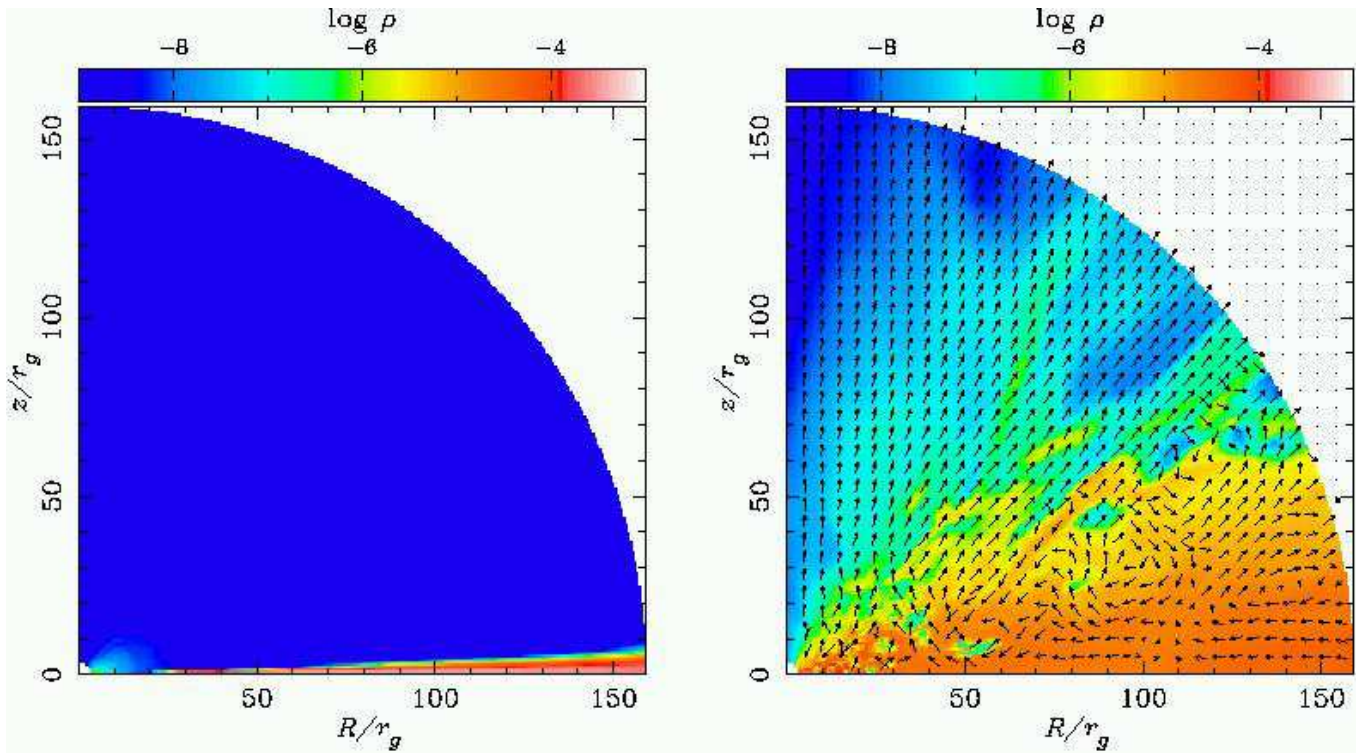}
\caption{
The 2D density distributions in the low-(left) 
and high-luminosity states (right).
The elapsed times are 166 s and 215 s, respectively.
The arrows in the right panel indicate the velocity vectors.
The adopted mass input rate and viscosity model are same as 
those in Figure 1.
}
\end{figure*}

In Figure 3, we plot the radiation temperature profiles 
at the equatorial plane (top panel)
and the effective temperature profiles (bottom panel)
in the high- and low-luminosity states.
Here, the radiation temperature is evaluated as
$T_{\rm r}\equiv (E_0/a)^{1/4}$, where $a$ is the radiation constant,
and it is roughly equivalent to the gas temperature 
around the equatorial plane. 
As shown in the top panel, 
the temperature of the disk is a few times larger 
in the high-luminosity state than in the low-luminosity state.
We calculate the effective temperature by solving 
the radiation transfer equation in the face-on view.
We find that the effective temperature is roughly proportional
to $r^{-3/4}$ in the low state and $r^{-1/2}$ in the high state
except for the very vicinity of the black hole
(see bottom panel).
They are consistent with the relations for 
the standard and slim disks.
\begin{figure}[b]
\epsscale{1.18}
\plotone{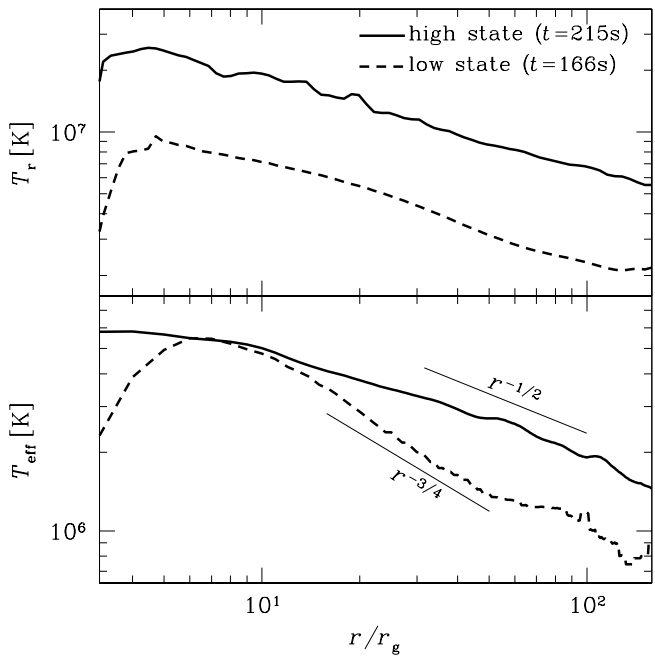}
\caption{
The radiation temperature profiles 
at the equatorial plane (top)
and the effective temperature profiles (bottom)
in the high- and low-luminosity states.
The adopted mass input rate and the viscosity model are 
the same as those in Figure 1.
}
\end{figure}

Here, we need to remark that the sum of the mass accretion rate
and outflow rate is about half of the mass input rate on average.
It means that the total mass within the computational domain 
monotonically increases with time.
However, the increase of the mass mainly occurs in 
the outer part of the disk ($r\gg 100r_g$),
and it does not influence the time evolution of the inner part of the disk
($r\lsim 100r_g$) so much.
Thus, we conclude that our simulations can reproduce 
the quasi-periodic oscillations of an unstable disk
around the black hole. 
Much longer computation time is required,
compared with the present simulations,
for reproducing $\dot{M}_{\rm input}=\dot{M}_{\rm acc}+\dot{M}_{\rm out}$
on average.

Finally, we prove that the limit-cycle behavior in our simulations
is triggered by the thermal instability 
in the radiation-pressure dominant region. 
According to the accretion disk theory,
the limit-cycle oscillations occur
when two conditions are satisfied.
One of them is that the 
mass accretion rate is moderately high, 
$\dot{M}_{\rm acc} \gsim L_{\rm E}/c^2$.
Another one is that the viscous stress tensor is
proportional to the total pressure.
To investigate the physical mechanism of the oscillations 
in our simulations,
we study whether or not the limit-cycle oscillations occur 
by assuming $\dot{M}_{\rm input}=10^3 L_{\rm E}/c^2$
(comparison model A) and 
$t_{r\varphi} \propto p_{\rm gas}$ (comparison model B).
The top panel in Figure 4 shows 
the time evolution of the mass accretion rate 
as well as the luminosity of the comparison model A.
The viscosity prescription is the same as that of the original model,
that is, $t_{r\varphi} \propto p_{\rm total}$ in the disk.
As shown in the panel, 
it is found that the disk becomes a quasi-steady
($\dot{M}_{\rm acc}\sim 100 L_{\rm E}/c^2$ and $L\sim 3 L_{\rm E}$).
This represents that the disk is stabilized 
when the mass accretion rate highly exceeds the critical value.
On the other hand, the disk with the viscosity model
of $t_{r\varphi} \propto p_{\rm gas}$
does also not exhibit the limit-cycle oscillations.
We plot the evolution of $\dot{M}_{\rm acc}$ and $L$ 
of the comparison model B in the bottom panel.
In this model, although the disk is not quasi-steady yet
($\dot{M}_{\rm input} \gg \dot{M}_{\rm acc}+\dot{M}_{\rm out}$),
the mass accretion rate is high enough to give rise to the instability
in the case of $t_{r\varphi} \propto p_{\rm total}$
(Chen and Taam 1993, see also Kato et al. 1998),
and the elapse time exceeds the viscous timescale
in the inner region,
$ t_{\rm vis}\sim 160 (M/10M_\odot)
(r/5r_g)^{3/2} (\alpha/0.1)^{-1} (H/0.01r)^{-2}$ s.
It means that the viscosity model of $t_{r\varphi} \propto p_{\rm gas}$ 
does not cause the disk instability.
Since we succeed in reproducing 
the prediction of the disk theory,
we can conclude that recurrent outbursts 
in our simulations are caused by the disk instability 
in the radiation-pressure dominant region.
\begin{figure}[t]
\epsscale{1.18}
\plotone{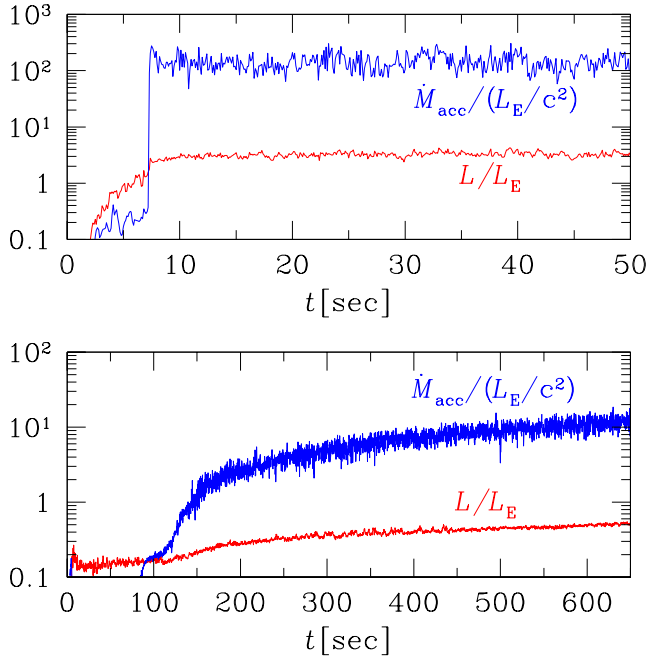}
\caption{
The time evolution of the mass accretion rate (blue) and 
the luminosity (red).
In the top panel, the viscosity model is the same as that in Figure 1.
But for high mass input rate, we adopt 
$\dot{M}_{\rm input}=10^3 L_{\rm E}/c^2$ (comparison model A).
On the other hand, the mass input rate is the same as that in Figure 1.
But for different viscosity law, 
we adopt $t_{r\varphi} \propto p_{\rm gas}$
(comparison model B).
}
\end{figure}
%
%



\section{DISCUSSION}
Resulting light curve in our simulations gives a nice fit to 
the time variation of the luminosity of GRS 1915+105.
Based on the analysis of the data of GRS 1915+105 
taken by RXTE and ASCA, Yamaoka et al. (2001) have produced the light curve
(see also Watarai \& Mineshige 2003).
The luminosity is several times higher in the high-luminosity state
than in the low-luminosity state.
The duration of the high-luminosity state is about $30-50$ s,
and there is sharp edges between the high
and low states.
Our numerical results succeed in reproducing these observed features
as shown in Figure 1.

In our simulations, the observed bursting behavior
is reproduced using the simple $\alpha$ prescription of the viscosity
proposed by Shakura \& Sunyaev (1973). 
[Note that our viscosity model 
gives $t_{r\varphi} \propto p_{\rm total}$ inside the disk
as we have already mentioned.]
This also might be purely 2D effect, 
since the modification of the viscosity model
is required by the 1D simulations.
Watarai \& Mineshige (2003) explained
the X-ray observations of GRS 1915+105
using 
$t_{r\varphi} \propto p_{\rm total} (p_{\rm gas}/p_{\rm total})^{0.1}$.
Some authors have proposed the phenomenological viscosity model,
in which the effects of the dissipation of the energy
in disk corona or in jets are simply taken into consideration
(Nayakshin et al. 2000; Janiuk et al. 2000, 2002, 2005),
to match the observed burst duration and luminosity.

The multi-dimensional motion, which is treated in our simulations,
would affect the time evolution of the disk. The mass, momentum/angular 
momentum, and energy are extracted from the disk by the outflow and 
transported by the convective motion and circulation.
Through these multi-dimensional effects,
the viscosity law might practically change and 
affect the limit-cycle behavior of the disk.
To investigate the role of the multi-dimensional motion
for the time evolution of the disk in detail is left as a future work.

The radial profiles of the temperature and pressure of the disk
in our (grid-based) simulations look similar to 
those of the SPH simulations (case 2 in Teresi et al. 2004b).
The temperature of the disk is a few times larger 
in the high-luminosity state than in the low-luminosity state
(see top panel in Figure 3),
and the radiation pressure exceeds the gas pressure 
by more than two orders of magnitude in the high state.
However, 
the duration time of the low-luminosity state 
in their SPH simulations is quite differ from our results. 
In their simulations, the disk stays in the low state for a few second
and it does not agree with observations of GRS 1915+105
(e.g. Janiuk \& Czerny 2005).
Our simulations indicate that the duration time 
of the low state is roughly comparable to that of the high state.
Also, the luminosity is a few times larger in their simulations 
than in our study, although 
the relatively small mass-input rate, $\sim 30 L_{\rm E}/c^2$,
is employed
in their simulations (we set $\dot{M}_{\rm input} = 100 L_{\rm E}/c^2$
in the present work).
These discrepancies may arise from the difference of the method of the 
numerical calculations.
Also, the treatment of the multi-dimensional effects might lead to
the difference of the time evolution of the disk.
As we have already mentioned, 
the outflow motion as well as the photon trapping
are carefully treated in our simulations,
although the SPH simulations by Teresi et al (2004b)
focused on the time evolution of the disk only.

Although we investigate the disk instability 
in the framework of the $\alpha$ prescription of the viscosity,
the realistic formalism about the viscosity should be investigated
from the magneto-hydrodynamical point of view
(e.g., Machida et al. 2000; Stone \& Pringle 2001).
We need radiation MHD simulations as future work.

\section{CONCLUSIONS}
By performing the 2D RHD simulations, 
we investigate the time evolution of the accretion disks
around the black holes and find the limit-cycle oscillations.
In this study, 
we carefully treat the 2D effects including the outflow and 
the photon trapping.
We summarize our results as follows:

(1)
The luminosity sharply rises 
from $\sim 0.3L_{\rm E}$ to $\sim 2L_{\rm E}$ in our simulations.
Duration of the high-luminosity state is about $30-50$ s.
The resulting variation amplitude and duration
nicely fit to the observations of microquasar, GRS 1915+105. 

(2)
It is also found that the 2D effects are significant 
in the high-luminosity state.
In this state, 
the trapped luminosity is comparable to the luminosity
due to the effective photon trapping.
The outflow, circular motion, and patchy structure
appear in this state.
The outflow is driven by the strong radiation force.

(3)
The physical mechanism, 
which causes the limit-cycle oscillations,
is the thermal instability 
in the radiation-pressure dominant region.
In our simulations, the disk is stabilized 
when the mass accretion rate highly exceeds the critical value.
The disk does not exhibit the oscillations
if the viscous stress tensor is proportional to the gas pressure only.

\acknowledgments

We would like to thank the anonymous reviewer for many helpful suggestions.
We especially thank K. Watarai, M. Mori, H. Susa, and N. Shibazaki
for useful comments and discussions.
The calculations were carried out 
by a parallel computer at Rikkyo University
and Institute of Natural Science, Senshu University.
%
We acknowledge Research Grant from Japan Society 
for the Promotion of Science (17740111).


\end{document}